\documentclass[12pt]{book}
\usepackage[dvips]{graphicx,color}
\usepackage{makeidx,tsukuba}
\makeauthorindex
\makeindex

\begin{document}

\BookTitle{\itshape The 28th International Cosmic Ray Conference}
\CopyRight{\copyright 2003 by Universal Academy Press, Inc.}
\pagenumbering{arabic}

\chapter{Detection of Tau Neutrinos in Underwater Neutrino Telescopes}
\author{
Edgar Bugaev,$^1$ Teresa Montaruli,$^{2,3}$ and Igor Sokalski$^{3,1}$\\[1mm]
{\it (1) INR, 60th October Anniversary Prospect 7a, 117312, Moscow, Russia\\
(2) Physics Department, Bari University, Via Amendola 173, 70126
Bari, Italy\\      
(3) INFN/Bari, Via Amendola 173, 70126 Bari, Italy\\
}}

\section*{Abstract}
If neutrinos produced in cosmological sources oscillate, neutrino telescopes 
can have a chance to detect signals from $\tau$-neutrinos. We present an 
estimate of the 'double bang' event rate  produced by $\nu_{\tau}$'s in km$^3$
scale detectors.

\section{Introduction}

$\nu_{\mu} \leftrightarrow \nu_{\tau}$ oscillations should lead to the 
proportion
$\nu_{e}\!\!:\!\!\nu_{\mu}\!\!:\!\!\nu_{\tau}\!=\!1\!\!:\!\!1\!\!:\!\!1$ 
for neutrinos produced in cosmological sources 
that reach the Earth, though the flavor ratio at 
production is expected to be  $1\!\!:\!\!2\!\!:\!\!0$ 
(if all muons decay). Passing through the Earth, $\nu_{\tau}$'s generate 
$\tau$-leptons via charged current (CC) interactions. Being short-lived 
particles, $\tau$'s decay in flight producing $\nu_{\tau}$'s and (in 
$\sim$35\% cases) also $\nu_{\mu}$'s or $\nu_{e}$'s. Hence, neutrinos of all 
flavors undergo a regeneration process. The Earth is transparent for VHE 
$\nu_{\tau}$'s and, moreover, $\nu_{\tau}$'s are able to recover partially 
also $\nu_{\mu,e}$ fluxes for which the Earth is opaque at VHE. $\nu_{\tau}$'s
can be detected via $\tau$'s produced in CC interactions. At energies 
$E_{\nu}<10^6$\,GeV $\tau$ range is short ($R_{\tau}<30$\,m) and showers at CC
interaction and $\tau$ decay vertices cannot be reconstructed separately in 
underwater/ice neutrino telescopes (UNT). Hence the 2 hadronic or 
hadronic+electromagnetic showers cannot be distinguished by a point-like
hadronic+electromagnetic shower, produced by a $\nu_e$. For 
$E_{\nu} \approx 10^{6\div7}$\,GeV the 'double bang' (DB) signature ($\tau$ 
track between 2 showers at $\nu_{\tau}$ interaction point and $\tau$ decay 
vertex [7]) can be a clear and background-free evidence of $\nu_{\tau}$ 
detection. At $E_{\nu}>3\cdot10^{7}$\,GeV $\tau$-lepton range becomes larger 
than 1\,km and $\tau$'s cannot be distinguished from muons in km$^3$ scale 
UNT. Using the muon and $\tau$ propagation code MUM~[13], we estimate that 
even though $\tau$'s and muons have different energy loss properties, a lower 
energy muon can produce through stochastic processes showers with similar 
features than a $\tau$-lepton.

\begin{figure}[h]
\begin{center}
\includegraphics[height=9.9cm]{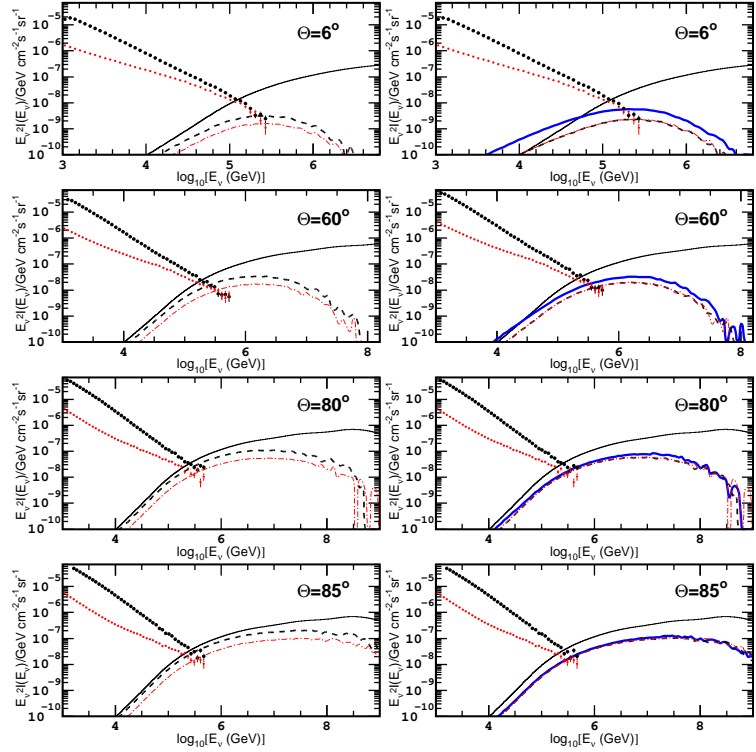}
\end{center}
\vspace{-1.0pc}
\caption{
Spectra of astrophysical $\nu$'s sampled according to the spectrum in [12] 
transformed after propagation through the Earth at 4 different nadir angles. 
Left column: no oscillations, 
$\nu_{e}\!\!:\!\!\nu_{\mu}\!\!:\!\!\nu_{\tau}\!=\!1\!\!:\!\!2\!\!:\!\!0$, 
right column: oscillations with maximal mixing, 
$\nu_{e}\!\!:\!\!\nu_{\mu}\!\!:\!\!\nu_{\tau}\!=\!1\!\!:\!\!1\!\!:\!\!1$. 
Thin solid lines: total incoming flux of astrophysical $\nu$'s; thick solid 
(right panels), dashed and dash-dotted lines (overlapping in right panels): 
out-coming $\nu_{\tau} + \bar \nu_{\tau}$, $\nu_{\mu} + \bar \nu_{\mu}$,
$\nu_{e} + \bar \nu_{e}$ after propagation through the Earth, respectively.
$\nu_{\mu,e}$ spectra include secondary neutrinos produced by $\nu_{\tau}$'s.
In the right part of each panel $\nu_{atm}$ spectra (conventional$+$prompt 
[10]) after propagation through the Earth are shown, upper thick markers: 
$\nu_{\mu} + \bar \nu_{\mu}$, lower thin markers: $\nu_{e} + \bar \nu_{e}$. 
}
\vspace{-5mm}
\end{figure}
\begin{figure}[t]
\begin{center}
\includegraphics[height=9.9cm]{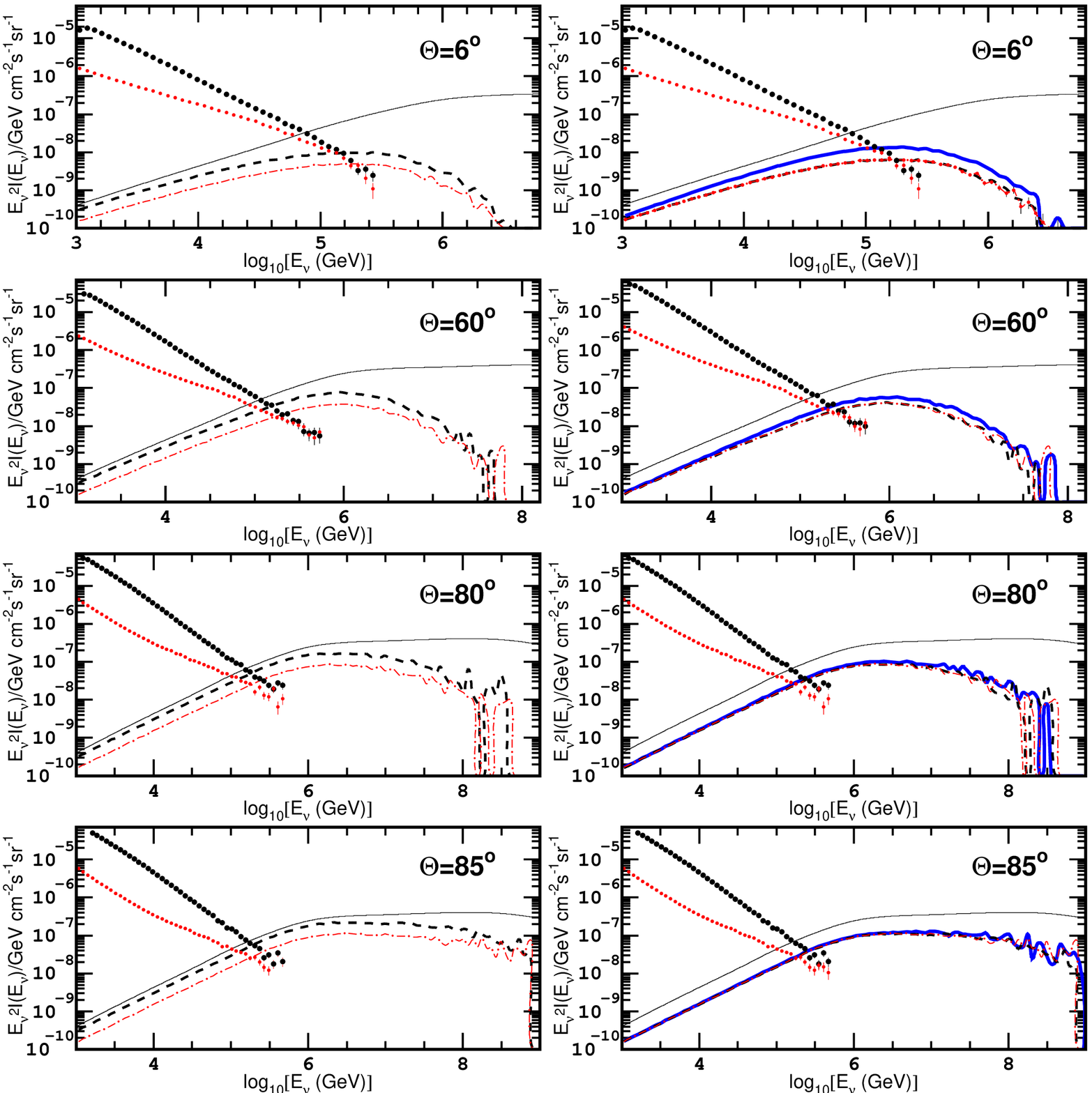}
\end{center}
\vspace{-1.0pc}
\caption{
The same as in Fig.\,1 but astrophysical neutrinos are generated according to 
Mannheim, Protheroe and Rachen upper limit on diffuse neutrino flux [9].
\vspace{-5mm}}
\end{figure}

Estimates of DB rates in km$^3$ UNT were done in [1] for down-going 
$\nu_{\tau}$'s, while events from the lower hemisphere were not considered.
Calculations of up-going $\nu_{\tau}$ propagation were done in [3]. It was 
concluded that, though VHE $\nu_{\tau}$'s are not absorbed by the Earth, their
energies decrease; hence the amount of $\tau$ events in the DB energy range is
low. In this work we present results of a MC simulation of astrophysical $\nu$ 
propagation through the Earth using 2 spectra [9,12] not considered in [1,3]. 
The rate of DB events in km$^3$ scale UNT is estimated.

\section{Method and Results}

We simulate both  DIS CC and NC interactions for all kind of neutrinos [8] in 
the energy range $10^3$\,GeV\,$<E_{\nu}<10^9$\,GeV using the CTEQ3\_DIS 
structure functions. MUM [13] was used for $\tau$ propagation including the 
newest corrections for photo-nuclear interaction [2]. We used TAUOLA package 
[6] to generate the $\tau$ decays. Comparison of results for power-law spectra
obtained by our algorithm with ones published in [4] showed a reasonable 
agreement. $\nu_{\mu,e}$ tracking is stopped after a CC interaction. $\tau$'s 
produced in $\nu_{\tau}$ CC interactions are propagated up to their decay 
vertex. Incident fluxes on Earth of $\nu$'s are assumed equal to $\bar \nu$ 
fluxes for all flavors. The background of atmospheric $(\pi\!K)$ and prompt 
(RQPM) $\nu_{\mu,e}$'s is generated using spectra in [10]. Simulation results 
are presented in Fig.\,1. We used the $\nu$ flux predicted by Protheroe [12] 
for an external photon optically thick proton blazar model. In Fig.\,2 
results
are presented or an upper bound (not for a model) on diffuse $\nu$ spectrum
for optically thick AGN sources (thick solid line in Fig.\,5 in [9]). 
$\nu_{\tau}$ fluxes exceed  $\nu_{\mu,e}$ ones remarkably up to nadir angles 
$\theta \sim 60^o$ since $\nu_{\tau}$'s are not absorbed by the Earth in 
contrast to $\nu_{\mu,e}$'s. But their spectra are shifted to lower energies 
due to energy degradation in regeneration process. For all $\theta$, the 
outgoing flux of astrophysical neutrinos exceeds $\nu_{atm}$ flux at 
$E_{\nu} > (1\div3)\cdot10^5$\,GeV. Secondary $\nu_{\mu,e}$'s which are 
produced by $\tau$ decays contribute to total outgoing $\nu_{\mu,e}$ spectra 
with fractions 0.57, 0.18, 0.06, 0.02 (spectrum [12]) and 0.18, 0.06, 0.02, 
0.01 (spectrum [9]) for nadir angles $\theta=$6$^o$, 60$^o$, 80$^o$, and 
85$^o$, respectively. Analysis of ratio between muon rates and shower rates 
that may be an indirect signature of $\nu_{\tau}$ appearance was presented in 
[3]. In this work we analyze direct $\nu_{\tau}$ detection through DB 
signature. The rate of totally contained DB events is given by:
\vspace{-1mm}
{\small
\begin{equation}
N=2\pi\,\rho\,N_A\int_{-1}^{0(1)}\!\!\int^{\infty}_{E_{min}}\!\!V_{eff}(E_{\nu_{\tau}},\theta)\,I(E_{\nu_{\tau}},\theta)\,\sigma^{CC}(E_{\nu_{\tau}})\,dE_{\nu_{\tau}}d(\cos\theta), 
\end{equation}
}

\vspace{-4mm}
\noindent
where $N_A$ is the Avogadro number, $\rho$ is medium density (we use
$\rho=$\,1\,g\,cm$^{-3}$ which is close to sea water/ice density), 
$I(E_{\nu_{\tau}},\theta)$ is the differential $\nu_{\tau}$ flux. The Earth 
shadowing effect is accounted for the lower hemisphere (upper limit of the 
integral 0, while $\cos\theta=1$ is used to compute the number of events for 
the whole sphere); $V_{eff}=S_{p}(\theta)\,(L-R_{\tau}(E_{\nu_{\tau}}))$, with
$S_p$ projected area for tracks generated 
isotropically in azimuth at
the fixed $\theta$ directions 
on a parallelepiped (IceCube-like 
$1\!\times\!1\!\times\!1$\,km$^3$ [5] and NEMO-like 
$1.4\!\times\!1.4\!\times\!0.6$\,km$^3$ [11]), $L$ geometrical distance 
between entry and exit point, $R_{\tau}$ $\tau$-lepton range, $\sigma^{CC}$  
total CC $\nu$ cross section. $E_{min}\!=\!2\cdot10^6$\,GeV corresponds to 
$\tau$-lepton range $R^{min}_{\tau}\!=\!70$\,m. DB can be considered as a 
background-free $\tau$ signature [7] since there should be no atmospheric 
$\mu$ which could produce 2 showers with comparable amount of photons. 
Nevertheless, this reasonable assumption needs to be verified through a full 
simulation. Table 1 shows DB rates for lower, upper and both hemispheres.
Values for upper hemisphere are 3$\div$6 times lower compared to [1] since 
more optimistic predictions for diffuse neutrino fluxes are used there. 
\vspace{-3mm}
\begin{table}[h]
\caption{Number of totally contained DB events in km$^3$ detector per year.}
\begin{center}
\begin{tabular}{c|cc}
\hline
Spectrum \rule{0pt}{4.0mm} \rule[-1mm]{0pt}{2.5mm} & IceCube-like ($N_{-2\pi}\,/\,N_{2\pi}\,/\,N_{4\pi}$) & NEMO-like ($N_{-2\pi}\,/\,N_{2\pi}\,/\,N_{4\pi}$)\\
\hline
$[12]$ \rule{0pt}{4.0mm} \rule[-1mm]{0pt}{2.5mm} & $0.7\,\,\,\,/\,\,\,\,1.6\,\,\,\,/\,\,\,\,2.3$      & $1.0\,\,\,\,/\,\,\,\,2.1\,\,\,\,/\,\,\,\,3.1$ \\
$[9]$  \rule{0pt}{4.0mm} \rule[-1.5mm]{0pt}{2.5mm} &  $1.0\,\,\,\,/\,\,\,\,2.3\,\,\,\,/\,\,\,\,3.3$      & $1.4\,\,\,\,/\,\,\,\,3.1\,\,\,\,/\,\,\,\,4.5$ \\
\hline
\end{tabular}
\end{center}
\end{table}

\vspace{-8mm}
\section{Conclusions}

Based on optimistic models [9,12] for AGN diffuse neutrino fluxes we found 
that in km$^3$ UNT one can expect a marginally observable rate of contained 
'double bang' $\tau$ events. 

\vspace{\baselineskip}

\re
1.\ Athar H.\ et al.\ 2000, Phys.\,Rev. D62, 093010 
\re
2.\ Bugaev E., Shlepin Yu.\ 2003, Phys.\,Rev. D67, 034027
\re
3.\ Dutta S. I.\ et al.\ 2000, Phys.\,Rev. D62 123001
\re
4.\ Dutta S. I.\ et al.\ 2002, Phys.\,Rev. D66 077302
\re
5.\ IceCube project: http://icecube.wisc.edu/
\re
6.\ Jadach S.\ et al.\ 1991, Comput.\,Phys.\,Commun. 64, 275
\re
7.\ Learned J. G., Pakvasa S.\ 1995, Astropart.\,Phys. 3, 267
\re
8.\ Generator that is based on cross sections described in Lipari P.\, et al.\ 1995, Phys.\,Rev.\,Lett. 74, 4384 with extension for VHE range by T.~Montaruli
\re
9.\ Mannheim K.\ et al.\ 2001, Phys.\,Rev. D63, 023003
\re
10.\ Naumov V. A.\ 2001, in Proc. of 2nd Workshop on Methodical Aspects of
Underwater/Underground Telescopes, Hamburg, p.31 [hep-ph/0201310]
\re
11.\ NEMO project: http://nemoweb.lns.infn.it/
\re
12.\ Protheroe R.\ 1997  in Accretion Phenomena and Related Outflows,
ed. Wickramasinghe et al. (Astronomical Society of the Pacific, San Francisco)
\re
13.\ Sokalski I.\ et al.\ 2001, Phys.\,Rev., D64, 074015
\endofpaper
\end{document}